\documentclass[conference]{IEEEtran}

\usepackage{amsmath}
\usepackage{amssymb}
\usepackage{amsfonts}
\usepackage{amsthm}
\usepackage{array}
\usepackage{cite}
\usepackage{float}

\newcommand{\bcomment}[1]{}
\newcommand{\TODO}[1]{}

\newcommand{\R}{\mathbb{R}}
\newcommand{\N}{\mathbb{N}}

\newcommand{\E}{\mathbb{E}}
\newcommand{\PP}{\mathbb{P}}

\newcommand{\diag}[1]{\operatorname{diag}(#1)}

\newcommand{\mcE}{\mathcal{E}}

\newcommand{\mcU}{\mathcal{U}}

\newcommand{\mcX}{\mathcal{X}}
\newcommand{\mcY}{\mathcal{Y}}
\newcommand{\mcZ}{\mathcal{Z}}

\newcommand{\eq}{\mathrel{\phantom{=}}}
\newcommand{\tr}{\operatorname{tr}}

\newcommand{\one}{\mathbf{1}}

\newcommand{\multisetds}[2]{\bigg(\kern-.4em\binom{#1}{#2}\kern-.4em\bigg)}
\newcommand{\multisetin}[2]{\big(\kern-.3em\binom{#1}{#2}\kern-.3em\big)}
\newcommand{\multisetix}[2]{\left(\kern-.2em\binom{#1}{#2}\kern-.2em\right)}

\newcommand{\eql}[1]{\overset{(#1)}{=}}
\newcommand{\leql}[1]{\overset{(#1)}{\leq}}

\DeclareMathOperator*{\argmax}{argmax}

\usepackage{color}
\usepackage[l2tabu,orthodox]{nag}
\usepackage{tikz}

\newtheorem{theorem}{Theorem}

\newtheorem{lemma}{Lemma}

\newtheorem{definition}{Definition}

\newcommand{\gf}{t}
\newcommand{\pr}{r}
\newcommand{\event}{\mcE}
\newcommand{\sumrv}{X}

\begin{document}
\title{Fundamental Limits of Database Alignment}
\author{\IEEEauthorblockN{Daniel Cullina}
\IEEEauthorblockA{Dept. of Electrical Engineering\\
Princeton University\\
dcullina@princeton.edu}
\and
\IEEEauthorblockN{Prateek Mittal}
\IEEEauthorblockA{Dept. of Electrical Engineering\\
Princeton University\\
pmittal@princeton.edu}
\and
\IEEEauthorblockN{Negar Kiyavash}
\IEEEauthorblockA{Dept. of Electrical and Computer Engineering\\
  Dept. of Industrial and Enterprise Systems Engineering\\
  Coordinated Science Lab\\
University of Illinois at Urbana-Champaign\\
kiyavash@illinois.edu}
}

\maketitle

\begin{abstract}
  We consider the problem of aligning a pair of databases with correlated entries.  
  We introduce a new measure of correlation in a joint distribution that we call cycle mutual information.
  This measure has operational significance: it determines whether exact recovery of the correspondence between database entries is possible for any algorithm.
  Additionally, there is an efficient algorithm for database alignment that achieves this information theoretic threshold.
\end{abstract}

\section{The database deanonymization problem}
Suppose that we have two databases.
Each item in the databases contain information about a single individual.
Some individuals appear in both databases.
When a entry in the first database and an entry in the second database concern the same individual, their contents are correlated.
The entries may be two noisy observations of the same signal, they may be two completely different types of data that have some correlation through population statistics, or they may even be correlated though the sampling process used to determine which individuals appear in the database.

We consider the following question:
If the databases are published with user identities removed from each entry, is it possible to learn the association between database entries that correspond to the same individual by exploiting the correlation between them?

Clearly, when there is enough correlation between entries about the same individual and the databases are small enough, it is possible to learn the true alignment between the database entries.
Our goal is to find the precise conditions under which it is possible to learn the complete correspondence between entries with high probability.
In particular, we would like to determine the measure of correlation that characterizes feasibility of perfect deanonymization in this setting.

This framework for database alignment is related to several practical deanonymization attacks.
Narayanan and Shmatikov linked an anonymized dataset of film ratings to a publicly available dataset using correlations between the ratings \cite{narayanan2008robust}.
Differential privacy has been widely used to quantifying privacy issues related to databases \cite{dwork2008differential}.
More recently, generative adversarial privacy has been proposed \cite{huang_context-aware_2017}.
In both cases, if users are present in multiple databases, knowledge of alignment is required to fully apply these frameworks.

Takbiri, Houmansadr, Goeckel, and Pishro-Nik have recently investigated a closely related user privacy problem \cite{takbiri_matching_2017}.

\subsection{Notation}
For finite sets $\mcX$ and $\mcY$, let $\R^{\mcX \times \mcY}$ be the set of real-valued matrices with rows indexed by $\mcX$ and columns indexed by $\mcY$.
For $x \in \R^{\mcX \times \mcY}$, let $x^{\odot k} \in \R^{\mcX \times \mcY}$ be the entry-wise power of $x$, i.e. the matrix such that $(x^{\odot k})_{i,j} = (x_{i,j})^k$.
Let $x^{\otimes k} \in \R^{\mcX^k \times \mcY^k}$ be the tensor power of $x$, i.e.
the matrix such that for $a \in \mcX^k$ and $b \in \mcY^k$, $(x^{\otimes k})_{a,b} = \prod_{i=0}^{k-1} x_{a_i,b_i}$.

Let $\PP(\mcX)$ be the set of probability distributions on $\mcX$.

\subsection{Formal description}
\begin{figure}
  \centering
  \begin{tikzpicture}[
  scale = 1/2,
  text height = 1.5ex,
  text depth =.1ex,
  b/.style={very thick}
  ]
  \draw (0,7) node {$M$};
  \draw (-6,7) node {$F_a$};
  \draw (6,7) node {$F_b$};
  \draw[dotted] (-2.5,6.5) -- (2.5,6.5) -- (2.5,1.5) -- (-2.5,1.5) -- cycle;
  \draw[dotted] (-7.5,6.5) -- (-3.5,6.5) -- (-3.5,1.5) -- (-7.5,1.5) -- cycle;
  \draw[dotted] ( 7.5,6.5) -- ( 3.5,6.5) -- ( 3.5,1.5) -- ( 7.5,1.5) -- cycle; 
  \draw (-3,7) node {$\mcU_a$};
  \draw ( 3,7) node {$\mcU_b$};
  \draw (-3,6) node {$u_1$};
  \draw (-3,5) node {$u_2$};
  \draw (-3,4) node {$u_3$};
  \draw (-3,3) node {$\vdots$};
  \draw (-3,2) node {$u_n$};

  \draw (3,6) node {$v_1$};
  \draw (3,5) node {$v_2$};
  \draw (3,4) node {$v_3$};
  \draw (3,3) node {$\vdots$};
  \draw (3,2) node {$v_n$};

  \draw[b] (-2.5,6) -- (2.5,4);
  \draw[b] (-2.5,5) -- (2.5,6);
  \draw[b] (-2.5,4) -- (2.5,3.1);
  \draw[b] (-2.5,3.1) -- (2.5,5);
  \draw[b] (-2.5,2.9) -- (2.5,2);
  \draw[b] (-2.5,2) -- (2.5,2.9);

  \draw (-6,6) node {$(0,1,1,1)$};
  \draw (-6,5) node {$(1,0,1,0)$};
  \draw (-6,4) node {$(0,1,0,0)$};
  \draw (-6,2) node {$(1,1,1,0)$};

  \draw (6,6) node {$(1,0,1,0)$};
  \draw (6,5) node {$(0,0,1,0)$};
  \draw (6,4) node {$(0,1,1,1)$};
  \draw (6,2) node {$(0,0,0,0)$};

  \draw[b,->] (-3.5,6) -- (-4.5,6);
  \draw[b,->] (-3.5,5) -- (-4.5,5);
  \draw[b,->] (-3.5,4) -- (-4.5,4);
  \draw[b,->] (-3.5,2) -- (-4.5,2);

  \draw[b,->] (3.5,6) -- (4.5,6);
  \draw[b,->] (3.5,5) -- (4.5,5);
  \draw[b,->] (3.5,4) -- (4.5,4);
  \draw[b,->] (3.5,2) -- (4.5,2);

\end{tikzpicture}
\caption{
  Two databases, $F_a$ and $F_b$, with alphabets $\mcX_a = \mcX_b = \{0,1\}^4$ and a matching $M$ between their user identifier sets.
}
\label{fig}
\end{figure}
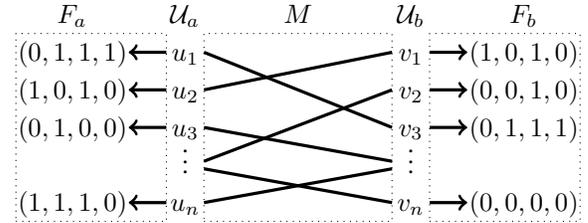

We have the following sets related to the user identifiers:
\vspace{-4mm}
\begin{center}
\begin{tabular}{ll}
  $\mcU_a$ & Set of user identifiers in the first database \\
  $\mcU_b$ & Set of user identifiers in the second database \\
  $M \subseteq \mcU_a \times \mcU_b$ & Bijective matching between the two types\\
           & of user identifierss \\
\end{tabular}
\end{center}
A bijection between $\mcU_a$ and $\mcU_b$ is a subset of $\mcU_a \times \mcU_b$ in which each element of $\mcU_a$ and $\mcU_b$ appears exactly once.
The matching $M$ contains the pairs of ids that correspond to the same user.
The fact that $M$ is a bijection implies that $|M| = |\mcU_a| = |\mcU_b|$.
Throughout, we let $n = |M|$.

We have the following sets, functions, and distributions associated with the databases:
\vspace{-4mm}
\begin{center}
\begin{tabular}{ll}
  $\mcX_a$ & Alphabet of entries in first database \\
  $\mcX_b$ & Alphabet of entries in second database \\
  $F_a : \mcU_a \to \mcX_a$ & First database \\
  $F_b : \mcU_b \to \mcX_b$ & Second database \\
  $F = (F_a,F_b)$\\
  $p \in \PP(\mcX_a \times \mcX_b)$ & Joint distribution between related entries\\
  $p_a \in \PP(\mcX_a)$ & Marginal distribution on first alphabet \\
  $p_b \in \PP(\mcX_b)$ & Marginal distribution on second alphabet \\
\end{tabular}
\end{center}
Figure~\ref{fig} illustrates a pair of databases.
\subsection{Generative model}
For each user $u \in \mcU_a$, there is a database entry $F_a(u) \in \mcX_a$.
For a pair $(u,v) \in M$, the entries $F_a(u)$ and $F_b(v)$ are correlated via the joint distribution $p$:
\[
  \Pr[F_a(u) = i, F_b(v) = j | M] = p(i,j) .
\]

For distinct $u,v \in \mcU_a$, $F_a(u)$ and $F_a(v)$ are independent.
The same is true for distinct $u,v \in \mcU_b$.
Thus we define
\[
  \pr(f_a,f_b;m) = \prod_{(u,v) \in m} p(f_a(u), f_b(v))
\]
so the joint distribution of the databases is
\begin{equation}
  \Pr[F_a = f_a, F_b = f_b | M = m] = \pr(f_a,f_b;m). \label{gen-model}
\end{equation}

\subsection{Relationship to graph alignment}
The methods used in this paper are related to those used to analyze information theoretic thresholds for exact graph alignment \cite{Cullina2017exact, cullina_improved_2016,pedarsani_privacy_2011}.
An undirected graph $G$ can be represented by its edge indicator function: $\binom{V(G)}{2} \to \{0,1\}$, so we have a very simple type of information about each user pair.
The analogue to the generative model \eqref{gen-model} is the correlated Erd\H{o}s-R\'{e}nyi distribution on graph pairs, where corresponding edge indicator r.v.s are sampled i.i.d. from some joint distribution on $\{0,1\}^2$.
Once the marginal distributions are fixed, the one remaining degree of freedom specifies the level of correlation.

In the database problem, we instead have larger blocks of information about individual users.
This allows for more complicated forms of correlations.
In this paper, we identify the relevant one-dimensional summary of that correlation.

A further connection is that graph alignment falls into the database alignment framework when seed vertices are used \cite{ji2015your,dai_performance_2018}: the list of adjacent seeds is essentially a database entry.

\section{Results}
\label{sec:results}
Both our achievability and converse bounds use the following measure of correlation in a joint distribution.
We propose to call this quantity \emph{cycle mutual information}.
\begin{definition}
  For $p \in \mathbb{P}(\mcX_a \times \mcX_b)$, let $z \in \mathbb{R}^{\mcX_a \times \mcX_b}$ be the matrix such that $z_{i,j} = \sqrt{p(i,j)}$ for $i \in \mcX_a$ and $j \in \mcX_b$.
  For an integer $\ell \geq 2$, define the order-$\ell$ cycle mutual information
  \[
    I^{\circ}_{\ell}(p) = \frac{1}{1-\ell} \log \tr ((zz^T)^{\ell}).
  \]
\end{definition}

Then $z$ has a singular value decomposition $z = U \Sigma V^T$ where $\Sigma = \diag{\sigma}$.
Observe that
\begin{equation}
  \tr(\Sigma^2) = \tr (U \Sigma V^T V \Sigma U^T) = \tr(zz^T) = \sum_{i,j} z_{i,j}^2 = 1, \label{b1is1}
\end{equation}
so $\sigma^{\odot 2}$, the vector of squared singular values, constitutes a probability distribution.
Thus we have another expression for cycle mutual information of order $\ell$:
\(
  I^{\circ}_{\ell}(p) = H_{\ell}(\sigma^{\odot 2}),
\)
where $H_{\ell}$ is the R\'{e}nyi entropy of order $\ell$.
This expression allows us to extend the definition of $I^{\circ}_{\ell}(p)$ to all nonnegative real $\ell$.

Our achievability theorem allows for arbitrary structure in the joint distribution of database entries.
\begin{theorem}
\label{thm:ach}
  Let $M \subseteq \mcU_a \times \mcU_b$ be a uniformly random bijection.
  Let the alphabets $\mcX_a$ and $\mcX_a$ and the joint distribution $p \in \mathbb{P}(\mcX_a \times \mcX_b)$ depend on $n$.
  If
  \[
    I^{\circ}_2(p) \geq 2 \log n + \omega(1) ,
  \]
  there is an estimator for $M$ given $F$ that is correct with probability $1-o(1)$.
\end{theorem}

When the database entries are vectors of independent identically distributed components, we have a converse bound with a leading term that matches the achievability.
\begin{theorem}
  \label{thm:converse}
  Let $M \subseteq \mcU_a \times \mcU_b$ be a uniformly random bijection.
  Fix alphabets $\mcY_a$ and $\mcY_b$ and a joint distribution $q \in \mathbb{P}(\mcY_a \times \mcY_b)$.
  Let $\mcX_a = \mcY_a^{\ell}$, $\mcX_b = \mcY_b^{\ell}$, and $p = q^{\otimes \ell}$, where $\ell$ can depend on $n$.
  If
  \[
    I^{\circ}_2(p) \leq (2 - \Omega(1))\log n ,
  \]
  any estimator for $M$ given $F$ is correct with probability $o(1)$.
\end{theorem}
\section{MAP estimation}
The optimal estimator for $M$ given $F$ is the maximum a posteriori estimator:
\begin{align*}
  \hat{m}(f_a,f_b)
  &= \argmax_m \Pr[M = m | F = (f_a,f_b)]\\
  &= \argmax_m \frac{\Pr[F = (f_a,f_b) | M = m] \Pr[M = m]}{\Pr[F = (f_a,f_b)]}\\
  &\eql{a} \argmax_m \Pr[F = (f_a,f_b) | M = m].
\end{align*}
In $(a)$ we use that fact that $M$ is uniformly distributed.

Define the event
\[
  \event_{m_2,m_1} = \{(f_a,f_b) : \pr(f_a,f_b;m_2) \geq \pr(f_a,f_b;m_1)\}.
\]
When $m_1$ is the true matching, this is the error event in which $m_2$ is incorrectly preferred to $m_1$.

\subsection{Algorithm for computing the MAP estimator}
Define the matrix $Q(f_a,f_b) \in \R^{\mcX_a \times \mcX_b}$,
\[
  Q(f_a,f_b)_{u,v} = \log p(f_a(u),f_b(v)).
\]
The MAP estimator is the max weight matching in $Q(f_a,f_b)$:
\[
  \hat{m}(f_a,f_b) = \argmax_m \sum_{(u,v) \in m} Q(f_a,f_b)_{u,v}.
\]
Thus $\hat{m}$ can be computed in $\mathcal{O}(n^3)$ time \cite{edmonds_theoretical_1972}.

\section{Generating functions}
Let $x$ and $y$ be two matrices of formal variables indexed by $\mcX_a \times \mcX_b$, and let $x_a$ and $y_a$ be vectors of formal variables indexed by $\mcX_a$, and let $x_b$ and $y_b$ be vectors of formal variables indexed by $\mcX_b$.
For a matching $m \in \mcU_a \times \mcU_b$ and a pair of databases $f_a : \mcU_a \to \mcX_a$ and $f_b : \mcU_b \to \mcX_b$,
define the generating function of the joint type
\begin{equation*}
  \gf(m;f_a,f_b;x) = \prod_{(u,v) \in m} x_{f_a(u), f_b(v)}.
\end{equation*}
Observe that $\gf(m;f_a,f_b;p) = \pr(f_a,f_b;m)$.

For a pair of matchings, define the generating function
\begin{multline*}
  B_{m_1, m_2}(x,y) = \sum_{f_a : \mcU_a \to \mcX_a} \sum_{f_b : \mcU_b \to \mcX_b}\\ \gf(m_1;f_a,f_b;x) \gf(m_2;f_a,f_b;y) .
\end{multline*}
By understanding the behavior of this generating function, we can obtains upper bounds on the probability of an estimator making an error.


Throughout this section, let $z \in \R^{\mcX_a \times \mcX_b}$ be a matrix and let $z_a \in \R^{\mcX_a}$ and $z_b \in \R^{\mcX_b}$ be vectors such that $z_{i,j} = \sqrt{p(i,j)}$, $(z_a)_i = \sqrt{p_a(i)}$, and $(z_b)_j = \sqrt{p_b(j)}$. 
\begin{lemma}
  \label{lemma:b-ub}
  For any two bijections $m_1,m_2 \subseteq \mcU_a \times \mcU_b$, 
  \[
    \Pr[\event_{m_2,m_1}|M = m_1] \leq B_{m_1, m_2}(z,z)
  \]
\end{lemma}
\begin{IEEEproof}
For any $\theta \geq 0$, we have 
\begin{align*}
  &\eq \Pr[\event_{m_2,m_1} | M = m_1]\\
  &=  \E\left.\left[\one\left(\frac{\pr(f_a,f_b;m_2)}{\pr(f_a,f_b;m_1)} \geq 1\right) \right| M = m_1\right]\\  
  &\leq \E\left.\left[\left(\frac{\pr(f_a,f_b;m_2)}{\pr(f_a,f_b;m_1)}\right)^{\theta} \right| M = m_1\right].
\end{align*}
Furthermore,
\begin{align*}
  &\eq \E\left.\left[\left(\frac{\pr(f_a,f_b;m_2)}{\pr(f_a,f_b;m_1)}\right)^{\theta} \right| M = m_1\right]\\
  &= \sum_{f_a,f_b} \left(\frac{\pr(f_a,f_b;m_2)}{\pr(f_a,f_b;m_1)}\right)^{\theta} \pr(f_a, f_b; m_1)\\
  &= \sum_{f_a,f_b} \pr(f_a, f_b; m_2)^{\theta} \pr(f_a, f_b; m_1)^{1-\theta}\\
  &= \sum_{f_a,f_b} \gf(m_1;f_a,f_b;p)^{\theta} \gf(m_2;f_a,f_b;p)^{1-\theta}\\
  &= \sum_{f_a,f_b} \gf(m_1;f_a,f_b;p^{\odot \theta}) \gf(m_2;f_a,f_b;p^{\odot (1-\theta)})\\
  &= B_{m_1,m_2}(p^{\odot \theta},p^{\odot (1-\theta)})
\end{align*}
where the matrix and vector exponents with $\odot$ are applied entrywise.
Selecting $\theta = \frac{1}{2}$ gives the claim.
\end{IEEEproof}

\bcomment{
\begin{lemma}
  \label{lemma:convex}
  Let
  \[
    c(\theta) = \log \sum_{f_a,f_b} \pr(f_a, f_b; m_2)^{\theta} \pr(f_a, f_b; m_1)^{1-\theta}
  \]
  Then $c(\theta)$ is convex.
\end{lemma}
\begin{IEEEproof}
  Let
  \begin{align*}
    \E\left.\left[\left(\frac{\pr(f_a,f_b;m_2)}{\pr(f_a,f_b;m_1)}\right)^{\theta} \right| M = m_1\right].\\
  \end{align*}
  \begin{align*}
    &\eq \exp(\gamma c(\theta) + (1-\gamma)c(\theta')) \\
    &= \gamma c(\theta) + (1-\gamma)c(\theta') \\
  \end{align*}
\end{IEEEproof}}

Define the generating function
\begin{align*}
  b^{\circ}_{\ell}(x,y) &= \tr((xy^T)^{\ell}).
\end{align*}

Regard $m_1$ as a function $\mcX_a \to \mcX_b$ and regard $m_2^T$ as a function $\mcX_b \to \mcX_a$.
Then their composition $m_2^T \circ m_1$ is a permutation of $\mcX_a$

\begin{lemma}
  \label{lemma:cycle-decomp}
  Let $m_1,m_2 \subseteq \mcU_a \times \mcU_b$ be bijections. 
  Let $t^{\circ}_{\ell}$ be the number of cycles of length $\ell$ in the permutation $m_2^T \circ m_1$.
  Then $t^{\circ}_1 = |m_1 \cap m_2|$, $\sum_{\ell} \ell t^{\circ}_{\ell} = |\mcX_a|$, and 
  \[
    B_{m_1,m_2}(x,y) = \prod_{\ell \in \N} (b^{\circ}_{\ell}(x,y))^{t^{\circ}_{\ell}}.
  \]
\end{lemma}

\begin{lemma}
  \label{lemma:norm}
  For $z' \in \R^{\mcX_a \times \mcX_b}$ with nonnegative entries and for $\ell \geq 2$, $b^{\circ}_{\ell}(z',z') \leq b^{\circ}_2(z',z')^{\ell/2}$.
\end{lemma}
\begin{IEEEproof}
We have $b^{\circ}_{\ell}(z',z') = \sum_k \sigma_k^{2\ell}$ where $\sigma_k$ are the singular values of $z'$.
By a standard inequality on $p$-norms, $\sum_k \sigma_k^{2\ell} \leq \left(\sum_k \sigma_k^4\right)^{\ell/2}$.
\end{IEEEproof}

\begin{lemma}
  \label{lemma:event-ub}
  Let $m_1,m_2 \subseteq \mcU_a \times \mcU_b$ be bijections and let $d = n - |m_1 \cap m_2|$.
  Then
  \[
    B_{m_1, m_2}(z,z) \leq b^{\circ}_2(z,z)^{d/2}.
  \]
\end{lemma}
\begin{IEEEproof}
  From \eqref{b1is1}, $b^{\circ}_1(z,z) = 1$.
  Then the claim follows from Lemmas~\ref{lemma:cycle-decomp} and \ref{lemma:norm}.
\end{IEEEproof}

\bcomment{
\begin{figure}
  \centering
  \begin{tikzpicture}[
  text height = 1.5ex,
  text depth =.1ex,
  b/.style={very thick}
  ]

  \definecolor{dred}{rgb}{0.7, 0.2, 0.2};

  \draw (-1,2) node {$\mcU_a$};
  \draw (-1,0) node {$\mcU_b$};

  \foreach \x in {0,1,2,3,4,5}
    \fill (\x,0) circle (2pt);

  \foreach \x in {0,1,2,3,4,5}
    \fill (\x,2) circle (2pt);

  \draw[->] (0,0) -- (0,2);
  \draw (1,0) -- (1,2);
  \draw (2,0) -- (2,2);
  \draw (3,0) -- (3,2);
  \draw (4,0) -- (4,2);
  \draw (5,0) -- (5,2);
  \draw[dred] (0,2) -- (6,2);
  
\end{tikzpicture}
\caption{Two matchings give a collection of cycles.}
\label{fig}
\end{figure}}
\section{Achievability}
\begin{IEEEproof}[Proof of Theorem~\ref{thm:ach}]
  We will use a union bound over all possible errors.
  \begin{align*}
    & \eq \Pr\bigg[\bigcup_{m_2 \neq m_1} \event_{m_2,m_1} \bigg| M = m_1 \bigg] \\
    & \leq \sum_{m_2 \neq m_1} \Pr[\event_{m_2,m_1}|M = m_1]\\
    & = \sum_{d=2}^n \sum_{m_2 \in S_{m_1,d}} \Pr[\event_{m_2,m_1}|M=m_1]
  \end{align*}
  where $S_{m,d}$ is the set of matchings that differ from $m$ is exactly $d$ places.
  We have
  \[
    |S_{m,d}| \leq \binom{n}{d} d! \leq n^d.
  \]
  From Lemma~\ref{lemma:b-ub} and Lemma~\ref{lemma:event-ub}, we have
  \begin{align*}
    \Pr[\event_{m_2,m_1}|M=m_1]
    &\leq \prod_{\ell} b^{\circ}_{\ell}(z,z)^{t^{\circ}_{\ell}}\\
    &\leq \prod_{\ell} (b^{\circ}_2(z,z)^{\ell/2})^{t^{\circ}_{\ell}}\\
    &= b^{\circ}_2(z,z)^{d/2} .
  \end{align*}  
  Thus the overall probability of error is at most
  \[
    \sum_{d=2}^n n^d b^{\circ}_2(z,z)^{d/2}.
\]
  From the main condition of the theorem, we have
  \begin{align*}
    I^{\circ}_2(p) &\geq 2 \log n + \omega(1)\\
    b^{\circ}_2(z,z) &\leq \exp(-2 \log n - \omega(1))\\
                   &= o(n^{-2}),
  \end{align*}
  so for sufficiently large $n$, $n b^{\circ}_2(z,z)^{1/2} < 1$ and we have 
  \begin{align*}
    \sum_{d=2}^n n^d b^{\circ}_2(z,z)^{d/2}
    \leq \frac{n^2 b^{\circ}_2(z,z)}{1 - n b^{\circ}_2(z,z)^{1/2}}
    \leq o(1)
  \end{align*}
  which proves the claim.  
\end{IEEEproof}

\section{Converse}

\begin{lemma}
  \label{lemma:sym}
  For any two bijections $m_1,m_2 \subseteq \mcU_a \times \mcU_b$, 
  \[
      B_{m_1, m_2}(x,y) = B_{m_2, m_1}(x,y).
  \]  
\end{lemma}
\begin{IEEEproof}
  For each $\ell$, $b^{\circ}_{\ell}(x,y) = b^{\circ}_{\ell}(y,x)$.
  The permutations $m_2^T \circ m_1$ and $m_1^T \circ m_2$ are inverses and thus have the same cycle decomposition.
  The claim follows from Lemma~\ref{lemma:cycle-decomp}.
\end{IEEEproof}

\begin{lemma}
  \label{lemma:event-lb}
    Fix alphabets $\mcY_a$ and $\mcY_b$ and a joint distribution $q \in \mathbb{P}(\mcY_a \times \mcY_b)$.
    Let $\ell$ depend on $n$ such that $\ell = \omega(1)$.
    Let $\mcX_a = \mcY_a^{\ell}$, $\mcX_b = \mcY_b^{\ell}$, $p = q^{\otimes \ell}$.
    For any two bijections $m_1,m_2 \subseteq \mcU_a \times \mcU_b$ such that $|m_1 \cap m_2| = n-2$, 
  \[
    \Pr[\event_{m_2,m_1}|M=m_1] \geq b^{\circ}_2(z,z)^{(1 + o(1))}.
  \]
\end{lemma}
\begin{IEEEproof}
  The function $c(\theta) = B_{m_1, m_2}(p^{\odot \theta},p^{\odot(1-\theta)})$ is a conditional moment generating function:
  \[
    c(\theta) = \E\left.\left[\exp\left(\theta\log \left(\frac{\pr(f_a,f_b;m_2)}{\pr(f_a,f_b;m_1)}\right)\right) \right| M = m_1\right] .
  \]  
  From Lemma~\ref{lemma:cycle-decomp}, we have
  \begin{align*}
    &\eq  B_{m_1, m_2}(p^{\odot \theta},p^{\odot(1-\theta)})\\
    &= b^{\circ}_1(p^{\odot \theta},p^{\odot(1-\theta)})^{n-2} b^{\circ}_2(p^{\odot \theta},p^{\odot(1-\theta)})\\
    &= b^{\circ}_2(p^{\odot \theta},p^{\odot(1-\theta)}) .
  \end{align*}
  because
  \[
    b^{\circ}_1(p^{\odot \theta},p^{\odot(1-\theta)}) = \tr((p^{\odot \theta})(p^{\odot (1-\theta)})^T) = \sum_{i,j} p_{i,j}^{\theta}p_{i,j}^{1-\theta} = 1.
  \]  
  By Lemma~\ref{lemma:sym}
  \[
    c(\theta) = b^{\circ}_2(p^{\odot \theta},p^{\odot(1-\theta)})=b^{\circ}_2(p^{\odot(1-\theta)},p^{\odot\theta}) = c(1-\theta).
  \]
  Moment generating functions are log-convex, so $c(\theta)$ is minimized at $\theta = \frac{1}{2}$.

  Because $p = q^{\otimes \ell}$, $c(\theta)$ is the product of $\ell$ identical terms.
  Let $u = q^{\odot \theta}$ and $v = q^{\odot (1-\theta)}$.
  \begin{align*}
    b^{\circ}_2(p^{\odot \theta},p^{\odot(1-\theta)})
    &= b^{\circ}_2(u^{\otimes \ell},v^{\otimes \ell})\\
    &= \tr((u^{\otimes \ell})(v^{\otimes \ell})^T(u^{\otimes \ell})(v^{\otimes \ell})^T)\\
    &= \tr(uv^Tuv^T)^{\ell}\\
    &= b^{\circ}_2(u,v)^{\ell}\\
    &= b^{\circ}_2(q^{\odot \theta},q^{\odot (1-\theta)})^{\ell}
  \end{align*}
  By Cram\'{e}r's Theorem on the asymptotic tightness of the Chernoff bound \cite{hajek_random_2015}
  \begin{align*}
    &\eq  \Pr\left[\log \left.\left(\frac{\pr(f_a,f_b;m_2)}{\pr(f_a,f_b;m_1)}\right) \geq 0 \right| M = m_1\right]\\
    &\geq b^{\circ}_2(q^{\odot \frac{1}{2}},q^{\odot \frac{1}{2}})^{\ell(1-o_{\ell}(1))}\\
    &=  b^{\circ}_2(p^{\odot \frac{1}{2}},p^{\odot \frac{1}{2}})^{1-o(1)}.
  \end{align*}
  Because $\ell = \omega(1)$, $o_{\ell}(1)$ and $o(1)$ are equivalent.
\end{IEEEproof}

\begin{lemma}
  \label{lemma:three-matchings}
  For any three bijections $m_1,m_2,m_3 \subseteq \mcU_a \times \mcU_b$,
  \[
    \Pr[\event_{m_2,m_1} \cap \event_{m_3,m_1} | M = m_1] \leq b^{\circ}_2(z,z)^{d/2}
  \]
  where $d = n - |m_2 \cap m_3|$.
\end{lemma}
\begin{IEEEproof}
  For $\theta \geq 0$ and $\theta' \geq 0$, 
\begin{align*}
  &\eq  \Pr\left.\left[
    \frac{\pr(f_a,f_b;m_2)}{\pr(f_a,f_b;m_1)} \geq 1 \wedge
    \frac{\pr(f_a,f_b;m_3)}{\pr(f_a,f_b;m_1)} \geq 1 \right| M = m_1\right]\\
  &= \E[\one(\event_{m_3,m_1}) \one(\event_{m_2,m_1}) | M = m_1 ]\\
  &\leq \E\left.\left[
    \left(\frac{\pr(f_a,f_b;m_2)}{\pr(f_a,f_b;m_1)}\right)^{\theta}
    \left(\frac{\pr(f_a,f_b;m_3)}{\pr(f_a,f_b;m_1)}\right)^{\theta'}\right| M = m_1\right]\\
  &= \sum_{f_a,f_b}
    \left(\frac{\pr(f_a,f_b;m_2)}{\pr(f_a,f_b;m_1)}\right)^{\theta}
    \left(\frac{\pr(f_a,f_b;m_3)}{\pr(f_a,f_b;m_1)}\right)^{\theta'}\pr(f_a, f_b; m_1)\\
  &= \sum_{f_a,f_b} \pr(f_a, f_b; m_2)^{\theta} \pr(f_a, f_b; m_3)^{\theta'} \pr(f_a, f_b; m_1)^{1-\theta - \theta'}
\end{align*}
Choosing $\theta = \theta' = \frac{1}{2}$, we obtain
\begin{align*}
  &\eq \E[\one(\event_{m_3,m_1}) \one(\event_{m_2,m_1}) | M = m_1 ]\\
  &\leq \sum_{f_a,f_b} \pr(f_a, f_b; m_2)^{\frac{1}{2}} \pr(f_a, f_b; m_3)^{\frac{1}{2}}\\
  &= B_{m_2,m_3}(z,z)\\
  &\leql{a} b^{\circ}_2(z,z)^{d/2}
\end{align*}
where $(a)$ follows from Lemma~\ref{lemma:event-ub}.
\end{IEEEproof}

\begin{IEEEproof}[Proof of Theorem~\ref{thm:converse}]
Let $m_1$ be the matching used to generate the databases and let $S = S_{m_1,2}$ be the set of matchings of size $n$ that differ from $m_1$ in exactly two places.
That is, for all $m \in S$, $|m_1 \cap m| = n-2$.
Observe that $|S| = \binom{n}{2}$, because each element of $S$ can be specified by the two users in $\mcU_a$ that it matches differently than $m_1$ does.
Let $\sumrv$ be the number of error events that occur:
\[
  \sumrv = \sum_{m \in S} \one(\event_{m,m_1}) .
\]
Let $\epsilon_1 = \Pr[\event_{m,m_1}|M = m_1]$, i.e. the probability that a specific transposition error occurs.

We need a lower bound on the probability that $\sumrv > 0$.
From Chebyshev's inequality, we have
\[
  \Pr\Big[(\sumrv - \E[\sumrv])^2 \geq \E[\sumrv]^2\Big] \leq \E\left[\frac{(\sumrv - \E[\sumrv])^2}{\E[\sumrv]^2}\right] = \frac{\E[\sumrv^2]}{\E[\sumrv]^2} - 1
\]
and we need to find conditions that make this $o(1)$.
We have
\begin{align*}
  \sumrv^2
  &= \sum_{(m_2,m_3) \in S^2} \one(\event_{m_2,m_1}) \one(\event_{m_3,m_1})\\
  &= \sum_{m_2 \in S} \one(\event_{m_2,m_1}) + 2 \sum_{\{m_2,m_3\} \in \binom{S}{2}} \one(\event_{m_2,m_1}) \one(\event_{m_3,m_1})
\end{align*}

\bcomment{Observe that
\begin{multline*}
  \binom{\binom{n}{2}}{2} = \frac{n(n-1)(n(n-1)-2)}{8} = \\
  \frac{n(n-1)(n+1)(n-2)}{8} = 3 \binom{n+1}{4} = 3 \binom{n}{4} + 3 \binom{n}{3}.
\end{multline*}
\[
  \binom{\binom{n}{2}}{2} = 3 \binom{n}{4} + 3 \binom{n}{3}.
\]}
For a set $\{m_2,m_3\} \in \binom{S}{2}$, either $|m_2 \cap m_3| = n-3$ or $|m_2 \cap m_3| = n-4$.
There are $3\binom{n}{3}$ pairs of the former type and $3\binom{n}{4}$ pairs of the latter type.
In the latter case, the indicator variables $\event_{m_2,m_1}$ and $\event_{m_3,m_1}$ are independent.
In the former case, let $\epsilon_2 = \Pr[\event_{m_2,m_1} \cap \event_{m_3,m_1}|M = m_1]$.

Now we compute
\begin{equation*}
  \E[\sumrv]^2 = \binom{n}{2}^2 \epsilon_1^2 = \left( \binom{n}{2} + 6 \binom{n}{3} + 6 \binom{n}{4} \right)\epsilon_1^2
\end{equation*}
and
\begin{align*}
  \E[\sumrv^2] &= \binom{n}{2} \epsilon_1 + 6 \binom{n}{3} \epsilon_2 + 6 \binom{n}{4} \epsilon_1^2\\
  \frac{\E[\sumrv^2] - \E[\sumrv]^2}{\E[\sumrv]^2}
  &= \frac{\binom{n}{2}(\epsilon_1 - \epsilon_1^2) + 6\binom{n}{3}(\epsilon_2 - \epsilon_1^2)}{\binom{n}{2}^2 \epsilon_1^2}\\
  &\leq \mathcal{O}\left(\frac{1}{n^2 \epsilon_1} + \frac{\epsilon_2}{n\epsilon_1^2}\right).
\end{align*}
From Lemma~\ref{lemma:three-matchings} we have $\epsilon_2 \leq (b^{\circ}_2(z,z))^{\frac{3}{2}}$ and from Lemma~\ref{lemma:event-lb} we have $\epsilon_1 \geq (b^{\circ}_2(z,z))^{1 + o(1)}$, so
\begin{equation*}
  \Pr[\sumrv = 0] \leq \mathcal{O}\left(\frac{1}{n^2 (b^{\circ}_2(z,z))^{1 + o(1)}} + \frac{1}{n(b^{\circ}_2(z,z))^{\frac{1}{2} + o(1)}}\right).
\end{equation*}
If $b^{\circ}_2(z,z) \geq n^{-2 + \Omega(1)}$, then
\[
  n^2 b^{\circ}_2(z,z)^{1+o(1)} \geq n^{2 + (1+o(1))(-2+\Omega(1))} \geq n^{\Omega(1)} \geq \omega(1)
\]
and $\Pr[X = 0] \leq o(1)$.
\end{IEEEproof}

\section{Properties of cycle mutual information}
Consider a joint distribution $p \in \PP(\mcX_a \times \mcX_b)$ and recall the definitions of $z$ and $\sigma$ from Section~\ref{sec:results}.
The properties of $\sigma^{\odot 2}$ reflect the correlation in the distribution $p$.
The following three conditions are equivalent: $\sigma^{\odot 2}$ is supported on one point, the rank of the matrix $z$ is one, and the $p$ is the product of distributions on $\mcX_a$ and $\mcX_b$.

$I^{\circ}_{\ell}(p)$ shares several properties with mutual information.
It is symmetric: $I^{\circ}_{\ell}(p) = I^{\circ}_{\ell}(p^T)$.
It tensorizes: $I^{\circ}_{\ell}(p^{\otimes k}) = k I^{\circ}_{\ell}(p)$.
It reduces to entropy in the case of identical random variables: if $\mcX_a = \mcX_b$ and $p = \diag{p'}$, then
\[
  I^{\circ}_{\ell}(\diag{p'}) = H_{\ell}(p').
\]
because $\sigma^{\odot 2}$ is a rearrangement of $p'$.
In general, we have 
\[
  I^{\circ}_{\ell}(p) \leq \min(H_{\ell}(p_a),H_{\ell}(p_b)).
\]
Something stronger is true: the distribution $\sigma^{\odot 2}$ majorizes $p_a$ and $p_b$.
The diagonal of $zz^T$ is the marginal distributions $p_a$:
\[
  (zz^T)_{i,i} = \sum_j  z_{i,j}^2 = \sum_j p_{i,j}.
\]
Furthermore,
\[
  (zz^T)_{i,i} = (U \Sigma V^T V \Sigma U^T)_{i,i} = \sum_k U_{i,k}^2 \sigma_k^2.
\]
Because $U$ is an orthogonal matrix, the Hadamard product $U \odot U$ is doubly stochastic.
Thus $\sigma^{\odot 2}$ majorizes $p_a$.
The diagonal of $z^Tz$ contains $p_b$, which is also majorized by $\sigma^{\odot 2}$.

\subsection{Data processing inequality}
\begin{lemma}
  Let $p \in \PP(\mcX)$, let $q \in \mcX \to \PP(\mcY)$, and let $r~\in~\mcY \to \PP(\mcZ)$,
  so $\diag{p} \in \PP(\mcX \times \mcX)$, $\diag{p}q \in \PP(\mcX \times \mcY)$, and $\diag{p}qr \in \PP(\mcX \times \mcZ)$.
  Then for integer $\ell \geq 2$, $I^{\circ}_{\ell}(\diag{p}q) \geq I^{\circ}_{\ell}(\diag{p}qr)$.
\end{lemma}
\begin{IEEEproof}
  Define the matrices $z_{i,k} = \sqrt{(\diag{p}q)_{i,k}}$ and $w_{i,l} = \sqrt{(\diag{p}qr)_{i,l}}$.
  Then 
  \[
    (zz^T)_{i,i} = (ww^T)_{i,i} = p_i
  \]
  We have
  \[
    (zz^T)_{i,j} = \sqrt{p_ip_j} \sum_{k \in \mcY} \sqrt{q_{i,k} q_{j,k}}.
  \]
  The sum is the Bhattacharyya coefficient of the distributions $q_{i,\cdot}$ and $q_{j,\cdot}$, which can be written in terms of the Bhattacharyya divergence as follows: $\exp\left(-\frac{1}{2}D_{\frac{1}{2}}(q_{i,\cdot}||q_{j,\cdot})\right)$.
  Similarly
  \[
    (ww^T)_{i,j} = \sqrt{p_ip_j} \sum_{l \in \mcZ} \sqrt{(qr)_{i,l} (qr)_{j,l}}.
  \]

  By the data processing inequality for R\'{e}nyi divergences \cite{van_erven_renyi_2014}, we have
  \[
    D_{\frac{1}{2}}(q_{i,\cdot}||q_{j,\cdot}) \geq D_{\frac{1}{2}}((qr)_{i,\cdot}||(qr)_{j,\cdot}).
  \]
  Thus
  \begin{align*}
    (zz^T)_{i,j} &\leq (ww^T)_{i,j}\\
    \tr((zz^T)^{\ell}) &\leq \tr((ww^T)^{\ell})\\
    I^{\circ}_{\ell}(\diag{p}q) &\geq I^{\circ}_{\ell}(\diag{p}qr)
  \end{align*}
  for all integer $\ell \geq 2$.
\end{IEEEproof}

\section*{Acknowledgement}
This work was supported in part by NSF grants CCF 16-19216, CCF 16-17286, and CNS 15-53437.
\bibliographystyle{IEEEtran}
\bibliography{IEEEabrv,deanon}

\end{document}